# Dielectric metasurface for independent complex-amplitude control of arbitrary two orthogonal states of polarization

Yanjun Bao[1,*], Ying Yu[2], Shang Sun[3], Yang Chen[3], Qing-Hua Xu[4] and Cheng-Wei Qiu[3,*]

[1]Institute of Nanophotonics, Jinan University, Guangzhou 511443, China

[2]School of Physics and Optoelectronic Engineering, Taiyuan University of Technology, Taiyuan 030024, China

[3]Department of Electrical and Computer Engineering, National University of Singapore, 4 Engineering Drive 3, Singapore 117583, Singapore

[4]Department of Chemistry, National University of Singapore, 4 Engineering Drive 3, Singapore 117583, Singapore

*Corresponding author. Email: yanjunbao@jnu.edu.cn (Y. B.); eleqc@nus.edu.sg (C.-W. Q.).

**Abstract:** Metasurfaces are planar structures that can manipulate the amplitude, phase and polarization (APP) of light at subwavelength scale. Although various functionalities have been proposed based on metasurface, a most general optical control, i.e., independent complex-amplitude (amplitude and phase) control of arbitrary two orthogonal states of polarizations, has not yet been realized. Such level of optical control can not only cover the various functionalities realized previously, but also enable new functionalities that are not feasible before. Here, we propose a single-layer dielectric metasurface to realize this goal and experimentally demonstrate several advanced functionalities, such as two independent full-color printing images under arbitrary elliptically orthogonal polarizations and dual sets of printing-hologram integrations. Our work opens the way for a wide range of applications in advanced image display, information encoding, and polarization optics.

# Introduction

Metasurface, two-dimensional planar structure composed of subwavelength scattered elements, provides a platform to manipulate the light at subwavelength scale, leading to a plethora of applications such as flat metalens [1-3], vortex beam generations [4-6] and holograms [7-9]. The most distinguishing feature of metasurface is its flexible ability to control the amplitude, phase and polarization (APP) of the light, based on which various functionalities have been realized. The light controls by different kinds of metasurfaces in the literatures are categorized in Fig. 1a. Due to the symmetry constraints, most of the existing designed metasurfaces are restricted to the two special polarizations, linear and circular ones. The metasurface elements can act as linearly birefringent devices with linear polarizations or impose additional geometric (Pancharatnam-Berry) phases [10] with circular polarizations. For these two polarizations, the designed metasurfaces in the literatures are mostly related to the amplitude [11-14] or phase manipulations [1-9], which represent a fundamental control of light. A higher level of light control is the independent manipulation of both amplitude and phase [15-19]. This enables more powerful and multifunctional control, such as printing-hologram integration [17, 19], improving hologram qualities [16, 18], et. al. Compared with the two special linear and circular polarizations, the manipulation of arbitrary polarization is more challenge. There are several works that have reported the arbitrarily control of the phase[20] or both the amplitude and phase[21] of the arbitrary polarization with suitable design.

The above works only address single polarization. As polarization has two eigen orthogonal states, it is necessary to control the two orthogonal polarizations independently for advanced functionalities. A typical structure is the one with elliptical or rectangular cross section. By altering its geometric sizes, the amplitude or phase of the two orthogonal linear polarizations along the long and short axes can be independently manipulated. The independent amplitude control can be used to generate two different printing images [22, 23] under the two orthogonal polarizations, which is applied for three-dimensional stereoscopic prints[23]. The independent phase control of two orthogonal polarizations has been reported for polarization controlled holograms [24, 25] and photonic spin splitting functionalities [26, 27]. However, these works are limited to linear or circular polarizations.

The independent control of arbitrary two orthogonal polarizations is an important issue for expanding the scope of metasurface optics. In 2017, Mueller et. al [28] demonstrated the independent control of arbitrary two orthogonal polarizations by combining the propagation phase and geometric phase. Recently, the independent amplitude control of arbitrary two orthogonal polarizations is also reported [29]. The highest level of optical control in Fig. 1a (indicated by red sphere), i.e., the independent amplitude and phase (or complex-amplitude) control of arbitrary two orthogonal polarizations, has not yet realized so far. Such control can not only cover all the realized functionalities mentioned above, but also make it possible to achieve advanced functionalities that are not feasible before.

Here, we propose a metasurface design to impose independent amplitude and phase profiles on a pair of arbitrary orthogonal polarizations in the output, as conceptually shown in Fig. 1b. For the proof of concept, we firstly demonstrate the independent amplitude control with the design of two full-color printing images under two arbitrary orthogonal polarizations. Next, the full independent complex-amplitude control of arbitrary orthogonal polarizations is verified by a dual sets of printing-hologram integration functionality, i.e., a metasurface can exhibit two printing images and two holographic images simultaneously.

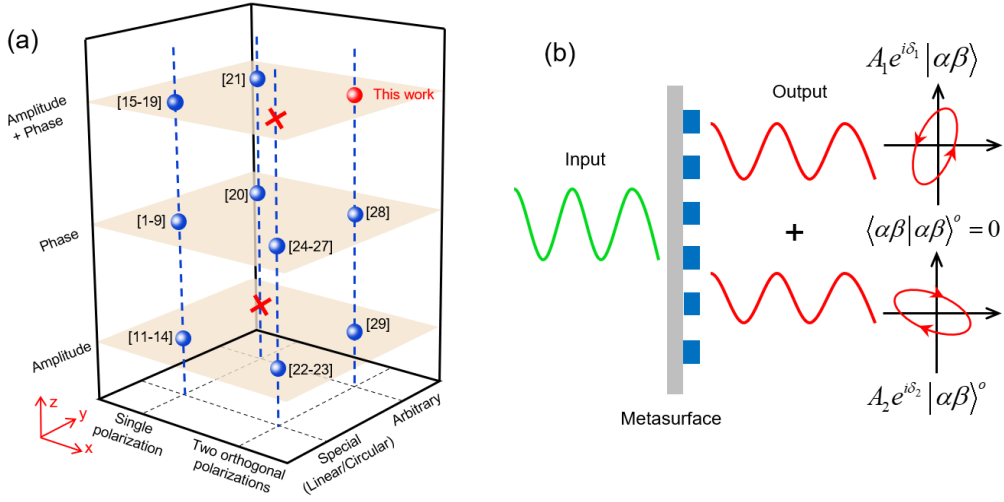

**Fig. 1 Metasurfaces for various light controls.** (a) A summary of the light control by metasurfaces in the literatures. The blue spheres and red crosses indicate the controls that have been realized and unrealized in the literatures, respectively. The reference numbers are illustrated beside the corresponding controls. The red sphere shows the light control of this work. The *xy* plane is the polarization controlled plane, including polarization numbers (one single polarization or two orthogonal polarizations) and polarization types (special or arbitrary). (b) Schematic of a metasurface for independent amplitude and phase control of arbitrary orthogonal polarizations in output. The two polarizations $|\alpha\beta\rangle$ and $|\alpha\beta\rangle^o$ are orthogonal with each other and have independent amplitudes $A_1$, $A_2$ and phases $\delta_1$, $\delta_2$, respectively.

## Results and Discussion

We start with two arbitrary orthogonal polarizations $|\alpha\beta\rangle = [\cos\alpha, \sin\alpha e^{i\beta}]^T$ and $|\alpha\beta\rangle^o = [\sin\alpha, -\cos\alpha e^{i\beta}]^T$ ($0 \leq \alpha \leq \pi/2$, $0 \leq \beta \leq 2\pi$) in the output (Fig. 1b). The aim of this work is to impose an independent amplitude $A_1$ and phase $\delta_1$ on an arbitrary polarization $|\alpha\beta\rangle$, and simultaneously impose an independent amplitude magnitude $A_2$ and phase $\delta_2$ on its orthogonal polarization $|\alpha\beta\rangle^o$. The previous works impose either independent amplitude[29] or phase[28], but here we can impose both. If the two optical fields coexist, the total output field can be written as $|T\rangle = A_1 e^{i\delta_1}|\alpha\beta\rangle + A_2 e^{i\delta_2}|\alpha\beta\rangle^o$. Such optical field has the property of maintaining their original designed amplitude and phase information under the selection of the corresponding polarization. For example, if the total output light is filtered with the allowance of only polarization $|\alpha\beta\rangle$, the complex-amplitude is $\langle\alpha\beta|T\rangle = A_1 e^{i\delta_1}$, which is the same as designed. This also applies for its orthogonal polarization $|\alpha\beta\rangle^o$. Therefore, filtered by a pair of arbitrary orthogonal polarizations, the output field can simultaneously generate two independent amplitude and two independent phase distributions into each eye of human, as an advanced version of traditional stereo displays.

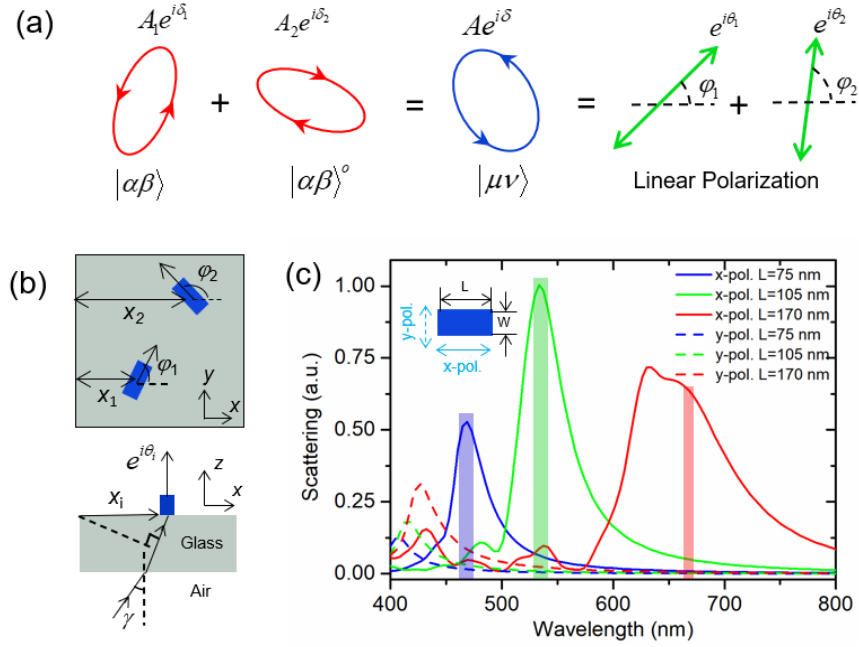

**Fig. 2 Metasurface design for independent complex-amplitude control of arbitrary two orthogonal polarizations.** (a) Decomposing the summation of two arbitrary orthogonal polarizations as the summation of two linear polarizations with pure phase modulations. The two orthogonal polarizations $|\alpha\beta\rangle$ and $|\alpha\beta\rangle^o$ have independent amplitudes $A_1$, $A_2$ and phases $\delta_1$, $\delta_2$, respectively. The general APP state has an amplitude $A$, phase $\delta$, and polarization $|\mu\nu\rangle$. (b) Schematic of the designed metasurface unit for representing the summation of two linear polarizations. The unit consists of two identical nanoblocks with strong anisotropy oriented at angle $\varphi_1$ and $\varphi_2$. The pure phase

modulations of the linear polarizations $\theta_1$, $\theta_2$ is acquired through detour phases under an incident angle of $\gamma$ in the $xz$ plane. (c) Simulated forward scattering intensity of a single c-silicon nanoblock with different lengths $L$ under $x$- (long axis) and $y$-polarized (short axis) incidences. The transparent color rectangles indicate the positions of incident wavelengths of the three RGB lasers.

The direct dealing with arbitrary polarizations is difficult, as most structures have only explicit optical response with linear or circular polarizations. Here, we discuss the possibility of the construction of the optical field with linear polarizations. As shown in Fig. 2a, the summation of the two orthogonal APP states $A_1 e^{i\delta_1}|\alpha\beta\rangle + A_2 e^{i\delta_2}|\alpha\beta\rangle^o$ results in a general APP state $Ae^{i\delta}|\mu\nu\rangle$ with amplitude $A$, phase $\delta$ and polarization $|\mu\nu\rangle$. Such optical state has four degrees of freedom (DOFs), and the final optical expression based on linear polarizations should have the same number of DOFs. We assume that it can be expressed as the summation of two linear polarizations with unit magnitude and pure phase modulations as (Fig. 2a):

$$A e^{i\delta}|\mu\nu\rangle = e^{i\theta_1}[\cos\varphi_1, \sin\varphi_1]^T + e^{i\theta_2}[\cos\varphi_2, \sin\varphi_2]^T \qquad (1)$$

where $\varphi_1$, $\varphi_2$ is the rotational angle of the two linear polarizations and $\theta_1$, $\theta_2$ is the pure phase modulations of each. If the analytical solutions of the parameters $\varphi_1$, $\varphi_2$, $\theta_1$ and $\theta_2$ can be obtained, the above assumption is true and actually it is. The detail of the derivation process of the analytical solutions is provide in Supplementary Material. Note that the decomposed two linear polarizations do not have to be orthogonal with each other.

For practical metasurface design, the linear polarization with rotational angle $\varphi_i$ (i=1, 2) can be constructed by a nanoblock oriented at the same angle $\varphi_i$, which should have strong anisotropy that only responds to the incident polarization along its long axis (Fig. 2b). The pure phase modulation can be introduced by the detour phase $\theta_i = 2\pi \sin\gamma x_i / \lambda$, where an incident beam with wavelength $\lambda$ is illuminated from the bottom of the metasurface at incident angle $\gamma$ and the scattered output field along normal direction (*z* axis) is considered (Fig. 2b). The detour phase of the nanoblock is proportional to its coordinate position $x_i$ and can be easily controlled. The commonly used propagation phase is not used here, as the phase can induce a non-controllable phase-dependent amplitude modulation.

Another issue is about the incident polarization. For different incident polarization, the component of electric field along the long axis of the nanoblock varies and therefore may add a non-unit amplitude modulation along with the detour phase. The pure phase modulation with unit amplitude can be only achieved with circular polarized light. In fact, it can be numerically verified that the general APP state $Ae^{i\delta}|\mu\nu\rangle$ can be decomposed as two linear polarizations with non-unit amplitude modulations for arbitrary incident polarization. In this work, we use the *y*-polarized incident polarization, as the incident electric field does not alter with the incident angle. For the *y*-polarized polarization, an amplitude modulation term $\cos\varphi_i$ is introduced in the two based linear polarizations. Although an analytical expression of $\varphi_1$, $\varphi_2$, $\theta_1$ and $\theta_2$ is difficult to obtain in this case, the values of these parameters can be solved numerically[21]. It is worth to mention that similar structure has been used for generating perfect vector beam[21], however, we use it here for a totally different purpose.

The nanoblock with strong anisotropy can be designed with suitable geometry parameters. For example, a crystal-silicon (c-silicon) nanoblock with length 170 nm, width 40 nm and height 600 nm can have an extinction ratio (defined as the ratio between the scattering intensities with incident polarization along the long axis and the short axis) over 50 in the wavelength range from 600 to 800 nm (Fig. 2c). Therefore, the scattering with the incident polarization along the short axis of the nanoblocks can be neglected. Furthermore, to cover full visible wavelength range, we introduce three nanoblocks with different geometry sizes that can best operate under red, green and blue (RGB) wavelengths as well as with minimal cross-talk between them. In experiment, we use three lasers at wavelengths of 671 nm, 532 nm and 473 nm to represent the RGB components. The optimized geometric sizes of the lengths of $R$, $G$ and $B$ nanoblocks are chosen as $L$=170 nm, 105 nm and 75 nm, respectively, and width $W$=40 nm and height $H$=600 nm for all ones. As shown in Fig. 2c, the simulated forward scattering spectra show a large extinction ratio for all the three different nanoblocks at their corresponding wavelengths, and therefore can be used as ideal building blocks for the concept verification of this work.

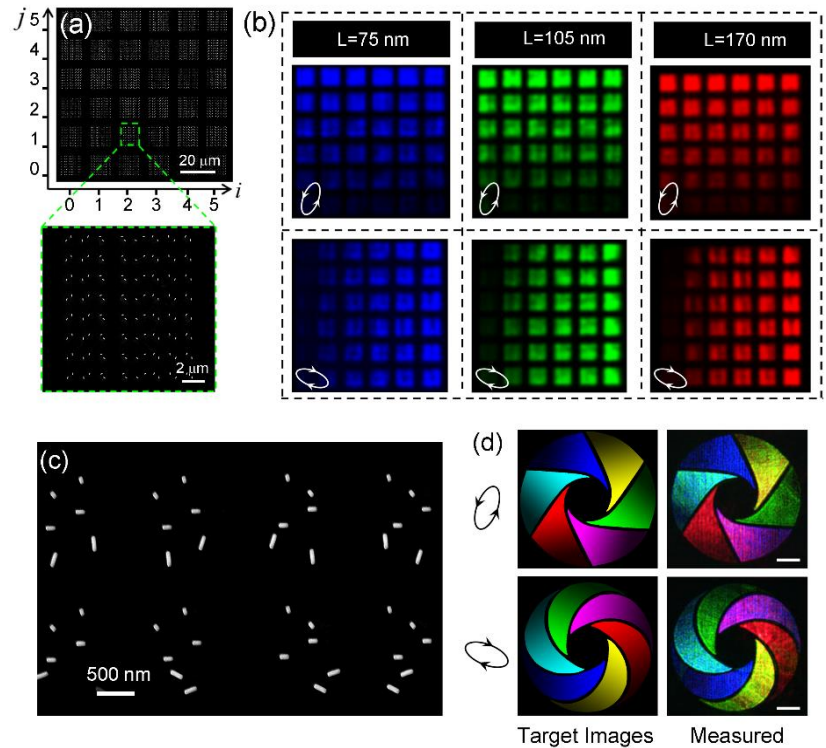

**Fig. 3 Experimental demonstration of full-color independent amplitude control of arbitrary two orthogonal polarizations.** (a) SEM images of a panel ($L$=170 nm) for verifying independent amplitude control of arbitrary orthogonal polarizations. The panel contains of 6×6 squares that possess varied intensity distributions under the two

orthogonal polarizations. (b) Measured optical images of the three different panels with the illuminations of 473 nm (first column), 532 nm (second column) and 671 nm (third column) lasers under the two orthogonal polarizations. The lengths of the nanoblocks are indicated in each column. (c) SEM images of the fabricated metasurfaces for independent two full-color printing images. The pixel of metasurface consists of three pairs of RGB double-nanoblock units and the period is set at 1.5 μm. (d) Target images (first column) and measured optical images (second column) under the two orthogonal polarizations selections. Scale bar: 50 μm.

We firstly give a proof-of-concept demonstration of full-color independent amplitude control of any pair of output orthogonal polarizations. The realization is based on the independent amplitude modulation of three types of nanoblocks under corresponding wavelengths of RGB. We consider a general case, a pair of elliptically orthogonal polarizations $|\alpha\beta\rangle$ and $|\alpha\beta\rangle^o$ with polarization parameters $\alpha=\pi/3$ and $\beta=\pi/3$. For each of the RGB wavelengths (671 nm, 532 nm and 473 nm), we design a panel containing 6×6 squares that possess different intensity distributions under the two orthogonal polarizations. In details, the square at coordinate $(i, j)$ ($i$=0, 1,···, 5, $j$=0, 1, ···, 5) is designed with an intensity of $I_{max}/5 \cdot i$ under one polarizations and an intensity of $I_{max}/5 \cdot j$ under its orthogonal one (Fig. 3a), where $I_{max}$ is the maximal intensity. The metasurface is made of c-silicon that is transferred on fused silica and fabricated using a standard electron beam lithography (see details of sample fabrication in Methods/Experimental). Figure 3a shows the scanning electron microscope (SEM) images of the metasurface panel ($L$=170 nm) designed for red wavelength.

To measure the varied intensities in the panels, the lasers are *y*-linearly polarized by a polarizer and obliquely incident on the metasurface. The scattered light is collected by a 4×/ 0.1 objective and filtered with a pair of quarter waveplate and polarizer for arbitrary selected polarization to pass through (see detail in Methods/Experimental). Figure 3b shows the measured optical images of three RGB panels (with different $L$ values) under the two orthogonal polarizations. In the first row, the intensities of squares for all the RGB panels increase along vertical direction but almost remain the same along the horizontal direction. For the corresponding orthogonal polarizations (second row), the situation is totally opposite. The measured results are consistent with our design, demonstrating the independent amplitude control of arbitrary two orthogonal polarizations. Besides the elliptical polarizations, we also verify the independent amplitude control of other orthogonal polarizations, including linear ($\alpha=0$, $\beta=0$) and circular ($\alpha=\pi/4$, $\beta=\pi/2$) cases (Supplementary Fig. S1).

By merging the three pairs of double-nanoblock RGB units into one pixel, independent two full-color printing images with varied intensities can be generated under the two orthogonal polarizations. Here, the period is purposely enlarged to give a clear visualization of each individual pixel (Fig. 3c). We choose two color wheels as the target images with their intensities gradually decreasing from inside to outside (Fig. 3d). Three lasers at wavelengths 671 nm, 532 nm and 473 nm are illuminated on the metasurface simultaneously to generate the full-color printing images. Figure 3d (second column) presents the measured optical images under the two orthogonal polarizations, which reproduce the varied intensities and colors of target images perfectly. The experimental results with each of the three lasers are provided in supplementary Fig. S2.

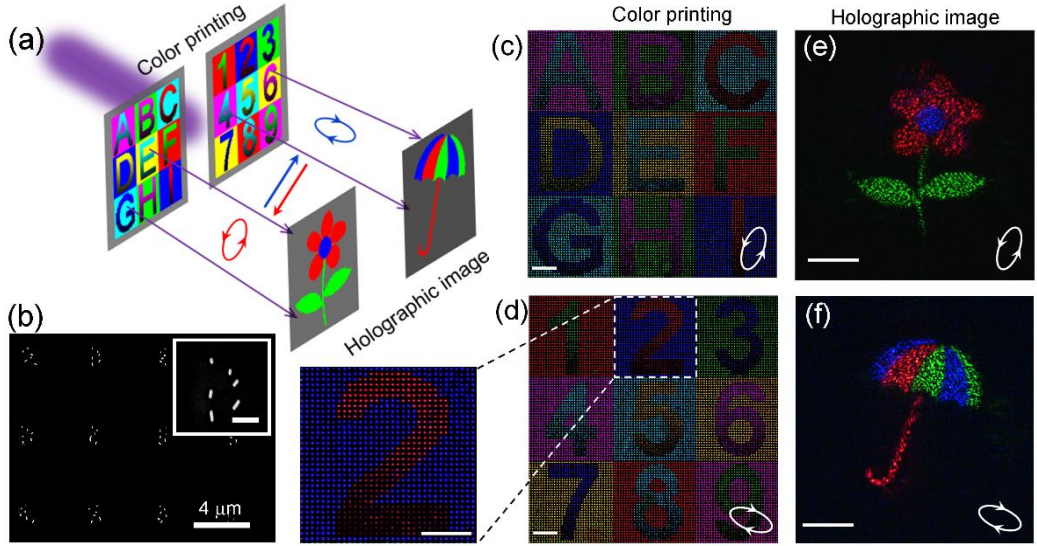

**Fig. 4 Experimental demonstration of independent complex-amplitude control of arbitrary two orthogonal polarizations.** (a) Schematic view of a metasurface that exhibits two sets of full-color printing and holography under arbitrary two orthogonal polarizations. The two printing images come from the same metasurface but are offset spatially for clarity. The holographic images are designed at a distance of 1.8 mm above the metasurface. The two orthogonal polarizations are designed as the same as that in Fig. 3d. (b) SEM images of the fabricated metasurface. The period is $P_x=P_y=5$ μm. The inset shows an enlarged view of one pixel (scale bar: 500 nm). (c-d) Measured optical printing images under the two orthogonal polarizations. An enlarged view of number 2 shows the continuously varied intensity. Scale bars: 50 μm. **e-f** Measured optical holographic images under the two orthogonal polarizations. Scale bars: 50 μm.

We then turn to verify the full independent complex-amplitude control of arbitrary orthogonal polarizations. Previously, the independent amplitude and phase control has been utilized for color-printing and holographic-image integration, which, however, operates only under one specific polarization of either circular [19] or linear [17] one. Therefore, the design strategy proposed in this work enables dual sets of printing and holographic images operated under arbitrary two orthogonal polarizations (Fig. 4a). Here, one thing that should be noted is that the phases of each metasurface pixels are not the same but determined by the hologram. If the period between adjacent pixels is too small, the intensities of the printing images can be disordered by the interferences of the adjacent pixels with different phases. One can design an enlarged pixel constructed by several sub-pixels with same phase [30] or increasing the period between pixels to avoid the interference. For a proof-of-concept verification, we simply increase the pixel pitch to 5 μm (Fig. 4b). The three pairs of RGB double-nanoblock units are merged into one pixel to realize full-color printing and holographic images (Fig. 4b). The algorithm to obtain the amplitude and phase distribution under each of the two orthogonal polarizations is the same as our previous work [19]. An objective (10×, 0.25 N.A.) is used to collect the scattered light by metasurface.

Figure 4a shows the two designed color printing images, in which the intensities of the numbers 1-9 in one image and the alphabet letters A-I in the other one vary continuously, with both having varied colors in the backgrounds. Figure 5c-d presents the measured printing images under the two orthogonal polarizations, which become pixelated due to the large separation between adjacent pixels. Aside from this difference, the measured printing images are almost identical to the designed ones, including the various colors and continuously varied brightness in the numbers and alphabet letters. For the two holographic images, we design a RGB flower and a RGB umbrella (Fig. 4a), which can be perfectly reproduced in the measurement (Fig. 4e-f). These results clearly demonstrate the independent complex-amplitude distributions imparted on two output orthogonal polarizations, which greatly expands the degree of freedom of light control in metasurface optics.

At last, it is worth to mention the difference between the previous works[28, 29] and ours. In our work, the independent amplitude and phase profiles are imposed on the two output orthogonal polarizations simultaneously with only one single incidence. The previous studies require the switching of the incidence polarizations to impose only amplitude[29] or phase[28] profiles on the two orthogonal outputs, which therefore cannot coexist simultaneously. Our strategy can be better used for some special situations, for example, stereo displays that require the simultaneous generation of two orthogonal polarizations with only one incidence. To realize the independent amplitude and phase control of arbitrary two orthogonal polarizations with switched incidence polarizations, the Jones matrix $J$ of the metasurface elements should have six DOFs, as $J = \begin{bmatrix} A_1 e^{i\gamma_1}, A_2 e^{i\gamma_2} \\ A_2 e^{i\gamma_2}, A_3 e^{i\gamma_3} \end{bmatrix}$, where $A_i$ and $\gamma_i$ (i=1, 2, 3) are the amplitude and phase tunings of each components of the Jones matrix (see Supplementary Materials). Such Jones matrix with six DOFs can be constructed with coherent pixel unit design, with the details discussed in our to-be published work[31].

**Conclusions**

In summary, we have demonstrated the independent complex-amplitude control of arbitrary two orthogonal polarizations, which represents the most general control of light. The core technique here is to decompose the summation of two arbitrary orthogonal polarizations as the summation of two linear polarizations, which then can be easily constructed with commonly used nanostructures (such as rectangle nanoblocks). With such control, we demonstrate new functionalized metasurface devices, including independent two full-color printing images under arbitrary orthogonal polarizations, and two sets of printing-holograms integration in a single metasurface structure. We also discuss the possible way to impose independent amplitude and phase with switched two incident orthogonal polarizations. We believe that our approach for the optical control of light could find diverse applications in future display system, imaging technology, high-density optical data storage, and complex field generation.

**Methods/Experimental**

*Sample fabrication.* The c-silicon on $SiO_2$ substrate is firstly prepared by transferring the thin-film c-silicon (1250 nm) from an SOI wafer onto $SiO_2$ substrate through adhesive wafer bonding and deep reactive ion etching (DRIE). Then inductively coupled plasma (ICP) is used to reduce the thickness of c-silicon layer to 600 nm. The fabrication of c-silicon metasurface is operated with electron beam lithography (EBL). In details, a 320 nm-thickness HSQ layer is firstly spin-coated at 4000 rpm on the c-silicon and then baked on a hot plate for 5 min at 90°C. Then a 30 nm-thickness aluminum layer (thermal evaporation) is deposited to serve as the charge dissipation layer. Next, the pattern is exposed using a Raith Vistec EBPG-5000plusES electron beam writer at 100 keV. After exposure, the aluminum layer is removed by 5% phosphoric acid and the resist is developed with tetramethylammonium hydroxide. Finally, the sample is etched using ICP.

*Optical measurement.* Three lasers at wavelengths of 671 nm, 532 nm and 473 nm are combined together and incident on the metasurface. The oblique incident angle is controlled by two mirrors and linearly polarized by a polarizer. The scattered light by the metasurface is collected by an objective and filtered with a pair of QWP and polarizer, which can be tuned to select the wanted polarization to pass through. A tube lens is used to focus the images on a CMOS color camera. A pinhole is inserted between the objective and sample to block the scattered stray lights. The optical setup for measuring the optical images of metasurface is shown in Supplementary Fig. S3.

**Declarations**

**Availability of data and materials**

The datasets used and/or analysed during the current study are available from the corresponding author on reasonable request.

**Competing interests**

The authors declare that they have no competing interests.

**Funding**

This work was supported by National Natural Science Foundation of China (62075246, 11804407). C.-W.Q. acknowledges the financial support by NUSRI via grant R-2018-S-001 and the support from the National Research Foundation, Prime Minister's Office, Singapore under its Competitive Research Programme (CRP award NRF CRP22-2019-0006). Y.C., Q.-H.X, and C.-W.Q. acknowledge financial support from Singapore MOR tier 2 Grant (R-143-000-A68-112).

**Authors' contributions**

Y. B. conceived the idea, performed the simulation, sample fabrication and experiment. Y.B. and C.-W.Q. wrote the manuscript. All authors discussed the results and commented on the manuscript.

**Acknowledgements**

Not applicable.

Supplementary Materials for

# Dielectric Metasurface for Independent Complex-Amplitude Control of Arbitrary Orthogonal States of Polarization

Yanjun Bao[1,*], Ying Yu[2], Shang Sun[3], Yang Chen[3], Qing-Hua Xu[4] and Cheng-Wei Qiu[3,*]

[1]Institute of Nanophotonics, Jinan University, Guangzhou 511443, China

[2]School of Physics and Optoelectronic Engineering, Taiyuan University of Technology, Taiyuan 030024, China

[3]Department of Electrical and Computer Engineering, National University of Singapore, 4 Engineering Drive 3, Singapore 117583, Singapore

[4]Department of Chemistry, National University of Singapore, 4 Engineering Drive 3, Singapore 117583, Singapore

*Corresponding author. Email: yanjunbao@jnu.edu.cn (Y. B.); eleqc@nus.edu.sg (C.-W. Q.).

# 1. Decomposing the summation of two arbitrary orthogonal polarizations as the summation of two linear polarizations

We consider two arbitrary orthogonal polarizations $|\alpha\beta\rangle = [\cos\alpha, \sin\alpha e^{i\beta}]^T$ and $|\alpha\beta\rangle^o = [\sin\alpha, -\cos\alpha e^{i\beta}]^T$ ($0 \leq \alpha \leq \pi/2$, $0 \leq \beta \leq 2\pi$) with independent amplitude $A_1$, $A_2$ and phase $\delta_1$, $\delta_2$, respectively. The summation of the two states is a general state with amplitude $A$, phase $\delta$ and polarization $|\mu\nu\rangle$, as

$$Ae^{i\delta}|\mu\nu\rangle = A_1 e^{i\delta_1}|\alpha\beta\rangle + A_2 e^{i\delta_2}|\alpha\beta\rangle^o \tag{S1}$$

The parameters of the general state can be obtained as

$$\begin{cases} A = \sqrt{A_1^2 + A_2^2} \\ \delta = \arg(A_1 e^{i\delta_1}\cos\alpha + A_2 e^{i\delta_2}\sin\alpha) \\ \mu = \arctan\left(\left|\dfrac{A_1 e^{i\delta_1}\sin\alpha - A_2 e^{i\delta_2}\cos\alpha}{A_1 e^{i\delta_1}\cos\alpha + A_2 e^{i\delta_2}\sin\alpha}\right|\right) \\ \nu = \arg\left(\dfrac{A_1 e^{i\delta_1}\sin\alpha - A_2 e^{i\delta_2}\cos\alpha}{A_1 e^{i\delta_1}\cos\alpha + A_2 e^{i\delta_2}\sin\alpha}\right) + \beta \end{cases} \tag{S2}$$

We then decompose it as a summation of two linear polarizations with pure phase modulation, as

$$Ae^{i\delta}|\mu\nu\rangle = e^{i\theta_1}[\cos\varphi_1, \sin\varphi_1]^T + e^{i\theta_2}[\cos\varphi_2, \sin\varphi_2]^T \tag{S3}$$

Equation S3 can be decomposed into the following two equations:

$$\begin{cases} Ae^{i\delta}\cos\mu = e^{i\theta_1}\cos\varphi_1 + e^{i\theta_2}\cos\varphi_2 & (S4) \\ Ae^{i\delta}\sin\mu e^{i\nu} = e^{i\theta_1}\sin\varphi_1 + e^{i\theta_2}\sin\varphi_2 & (S5) \end{cases}$$

By multiplying Eq. S5 by an imaginary number, and adding or subtracting it with Eq. S4, we obtain:

$$\begin{cases} Ae^{i\delta}[\cos\mu + i\sin\mu e^{i\nu}] = e^{i(\theta_1+\varphi_1)} + e^{i(\theta_2+\varphi_2)} & (S6) \\ Ae^{i\delta}[\cos\mu - i\sin\mu e^{i\nu}] = e^{i(\theta_1-\varphi_1)} + e^{i(\theta_2-\varphi_2)} & (S7) \end{cases}$$

The terms at the left-hand side can be written as:

$$\begin{cases} \cos\mu + i\sin\mu e^{iv} = \sqrt{1-\sin 2\mu \sin v}\, e^{i\gamma_1} = B_1 e^{i\gamma_1} & (S8) \\ \cos\mu - i\sin\mu e^{iv} = \sqrt{1+\sin 2\mu \sin v}\, e^{i\gamma_2} = B_2 e^{i\gamma_2} & (S9) \end{cases}$$

where $\gamma_1 = \mathrm{atan2}(\sin\mu\cos v, \cos\mu - \sin\mu\sin v)$, $\gamma_2 = \mathrm{atan2}(-\sin\mu\cos v, \cos\mu + \sin\mu\sin v)$ and the function atan2(y, x) returns the principal value of the arc tangent of y/x, expressed in radians. The range of the amplitude $B_1$ and $B_2$ is $0 \le B_{1,2} \le \sqrt{2}$. Then Eq. S6-7 can be rewritten as

$$\begin{cases} AB_1 e^{i(\delta+\gamma_1)} = e^{i(\theta_1+\varphi_1)} + e^{i(\theta_2+\varphi_2)} = 2\cos\left(\frac{\theta_1-\theta_2+\varphi_1-\varphi_2}{2}\right) e^{i(\theta_1+\theta_2+\varphi_1+\varphi_2)/2} & (S10) \\ AB_2 e^{i(\delta+\gamma_2)} = e^{i(\theta_1-\varphi_1)} + e^{i(\theta_2-\varphi_2)} = 2\cos\left(\frac{\theta_1-\theta_2-\varphi_1+\varphi_2}{2}\right) e^{i(\theta_1+\theta_2-\varphi_1-\varphi_2)/2} & (S11) \end{cases}$$

The above equations have two groups of solutions as

$$\begin{cases} \theta_1 = \delta + \frac{\gamma_1+\gamma_2}{2} + \frac{\mathrm{acos}(AB_1/2) \pm \mathrm{acos}(AB_2/2)}{2} & (S12) \\ \theta_2 = \delta + \frac{\gamma_1+\gamma_2}{2} - \frac{\mathrm{acos}(AB_1/2) \pm \mathrm{acos}(AB_2/2)}{2} & (S13) \\ \varphi_1 = \frac{\gamma_1-\gamma_2}{2} + \frac{\mathrm{acos}(AB_1/2) \mp \mathrm{acos}(AB_2/2)}{2} & (S14) \\ \varphi_2 = \frac{\gamma_1-\gamma_2}{2} - \frac{\mathrm{acos}(AB_1/2) \mp \mathrm{acos}(AB_2/2)}{2} & (S15) \end{cases}$$

Therefore, the summation of two arbitrary orthogonal polarizations can be considered as the summation of two linear polarizations with pure phase modulations.

## 2. Independent amplitude and phase control of two arbitrary orthogonal polarizations with switched incidences

Here, we discuss the case with two orthogonal incident polarizations, which differs with our work with a single incidence. If an incident polarization $|\alpha_1\beta_1\rangle$ is incident on the metasurface and an amplitude $A_1$ and phase $\delta_1$ is imposed on the output polarization

$|\alpha_2\beta_2\rangle$, the Jones matrix of the metasurface satisfies

$$A_1 e^{i\delta_1} |\alpha_2\beta_2\rangle = J|\alpha_1\beta_1\rangle \tag{S16}$$

In addition, for the orthogonal incident polarization $|\alpha_1\beta_1\rangle^o$, the metasurface also imposes an amplitude $A_2$ and phase $\delta_2$ on the output polarization $|\alpha_2\beta_2\rangle^o$. Therefore, the Jones matrix of the metasurface also satisfies

$$A_2 e^{i\delta_2} |\alpha_2\beta_2\rangle^o = J|\alpha_1\beta_1\rangle^o \tag{S17}$$

Note that $\langle \alpha_1\beta_1 | \alpha_1\beta_1 \rangle^o = 0$ and $\langle \alpha_2\beta_2 | \alpha_2\beta_2 \rangle^o = 0$. Considering that the incident polarization and output polarization is $|\alpha_2\beta_2\rangle = |\alpha_1\beta_1\rangle^*$, where * denotes the complex conjugate, we can express the above polarizations in the linear polarizations as

$$\begin{aligned}
|\alpha_1\beta_1\rangle &= [\cos\alpha_1, \sin\alpha_1 e^{i\beta_1}]^T \\
|\alpha_1\beta_1\rangle^o &= [\sin\alpha_1, -\cos\alpha_1 e^{i\beta_1}]^T \\
|\alpha_2\beta_2\rangle &= [\cos\alpha_1, \sin\alpha_1 e^{-i\beta_1}]^T \\
|\alpha_2\beta_2\rangle^o &= [\sin\alpha_1, -\cos\alpha_1 e^{-i\beta_1}]^T
\end{aligned} \tag{S18}$$

Then the Jones matrix $J$ is

$$\begin{aligned}
J &= \begin{bmatrix} A_1 e^{i\delta_1} \cos\alpha_1 & A_2 e^{i\delta_2} \sin\alpha_1 \\ A_1 e^{i\delta_1} \sin\alpha_1 e^{-i\beta_1} & -A_2 e^{i\delta_2} \cos\alpha_1 e^{-i\beta_1} \end{bmatrix} \begin{bmatrix} \cos\alpha_1 & \sin\alpha_1 \\ \sin\alpha_1 e^{i\beta_1} & -\cos\alpha_1 e^{i\beta_1} \end{bmatrix}^{-1} \\
&= \begin{bmatrix} A_1 e^{i\delta_1} \cos^2\alpha_1 + A_2 e^{i\delta_2} \sin^2\alpha_1, & \cos\alpha_1 \sin\alpha_1 e^{-i\beta_1}(A_1 e^{i\delta_1} - A_2 e^{i\delta_2}) \\ \cos\alpha_1 \sin\alpha_1 e^{-i\beta_1}(A_1 e^{i\delta_1} - A_2 e^{i\delta_2}), & e^{-i2\beta_1}(A_1 e^{i\delta_1} \sin^2\alpha_1 + A_2 e^{i\delta_2} \cos^2\alpha_1) \end{bmatrix}
\end{aligned} \tag{S19}$$

It can be observed that the Jones matrix is symmetric ($J_{12}=J_{21}$) and has a maximal six degrees of freedom. Therefore, if a Jones matrix with such six degrees of freedom as

$$J = \begin{bmatrix} A_1 e^{i\gamma_1}, & A_2 e^{i\gamma_2} \\ A_2 e^{i\gamma_2}, & A_3 e^{i\gamma_3} \end{bmatrix}$$

can be constructed, it can impose independent amplitude and phase on the two orthogonal polarizations of the incident light.

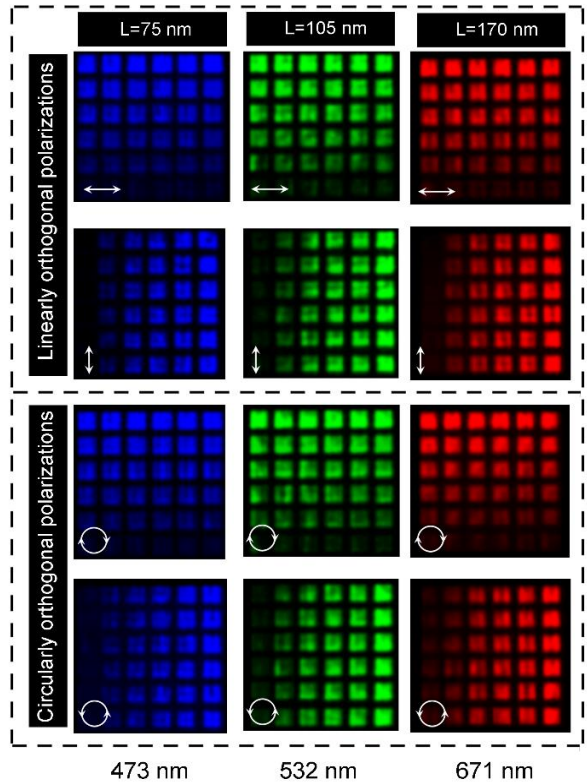

Fig. S1. Measured optical images of the designed panels under linearly and circularly orthogonal polarizations at three RGB incident wavelengths (473 nm, 532 nm, 671 nm).

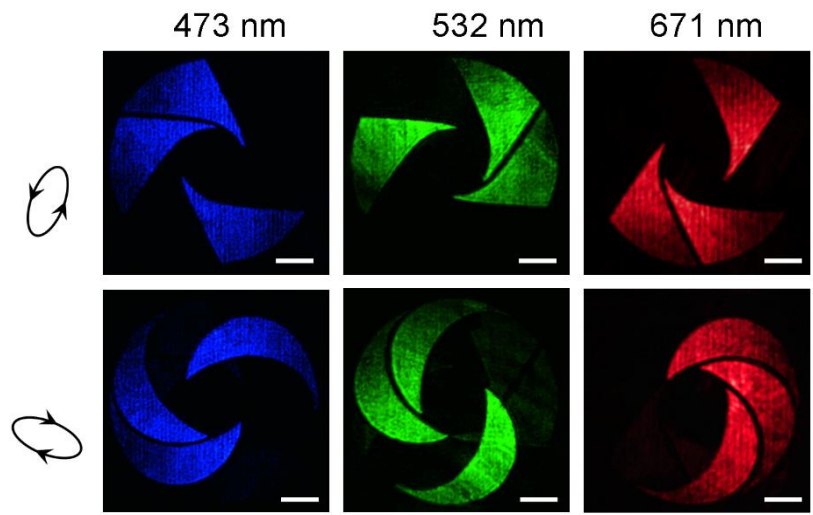

Fig. S2. Measured optical images when each of the three RGB lasers are illuminated individually (B, third column, G, fourth column, R, fifth column) under the two orthogonal polarizations. The wavelengths of the lasers are indicated in each panel.

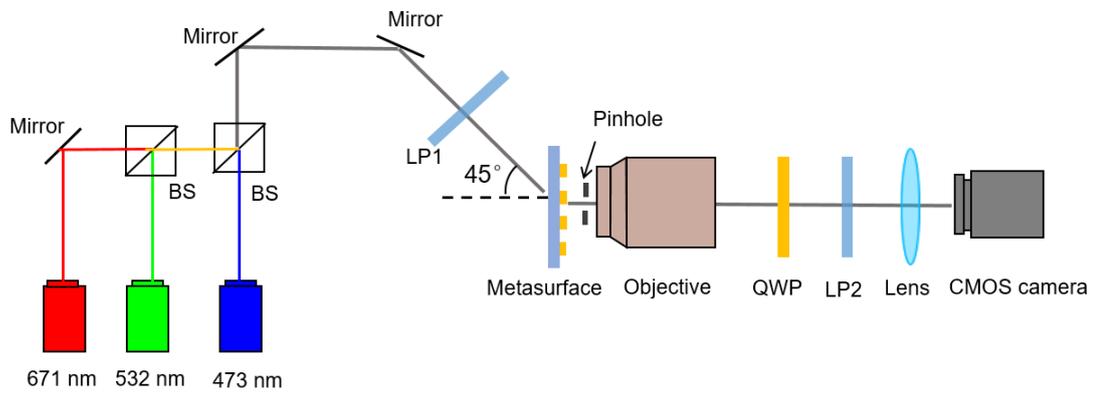

Fig. S3. Optical setup for measuring the printing images and holographic images. BS: beam splitter, LP: polarizer, QWP: quarter waveplate.